\documentclass[11pt]{article}
\usepackage{graphicx}

\newcommand{\BABARPubYear}    {02}

\newcommand{\BABARConfNumber} {024}
\newcommand{\SLACPubNumber} {9319}

\input pubboard/babarsym
\def\lumi    {84 million}
\def\limitrhoo {\ensuremath 1.4\times 10^{-6}}
\def\limitrhop {\ensuremath 2.3\times 10^{-6}}
\def\limitomega {\ensuremath 1.2\times 10^{-6}}
\def\limitgeneric {\ensuremath 1.9\times 10^{-6}}
\def\onlumi {\ensuremath 78}
\def\limitratio {\ensuremath 0.047}

\def\Bp {B^+}

\def\Bpm{B^{\pm}}
\def\bsg    {\ensuremath {b \to s \gamma}}
\def\bdg    {\ensuremath {b \to d \gamma}}
\def\bkg    {\ensuremath {\B \to \Kstar \gamma}}
\def\bkog    {\ensuremath {\Bz \to \Kstarz \gamma}}

\def\bkpg    {\ensuremath {\Bp \to \Kstarp \gamma}}

\def\brg    {\ensuremath {B \to \rho \gamma}}
\def\brog    {\ensuremath {\Bz \to \rho^0 \gamma}}
\def\brpg    {\ensuremath {\Bu \to \rho^+ \gamma}}
\def\bomg    {\ensuremath {\Bz \to \omega \gamma}}

\def\Kz    {\ensuremath{K^{0}}\xspace} 

\def\tg     {\ensuremath {\theta^{*}_T}}
\def\ctt     {\ensuremath {\cos{\tg}}}
\def\cth     {\ensuremath {\cos{\theta_{H}}}}
\def\ctb     {\ensuremath {\cos{\theta^{*}_{B}}}}

\def\de        {\ensuremath {\Delta E^{*}}}

\def\mpl #1 #2 #3 {Mod.~Phys.~Lett.~{\bf#1},\ #2 (#3)}
\def\npb  #1 #2 #3 {Nucl.~Phys.~B~{\bf#1},\ #2 (#3)}
\def\plb  #1 #2 #3 {Phys.~Lett.~B~{\bf#1},\ #2 (#3)}
\def\pr   #1 #2 #3 {Phys.~Rep.~{\bf#1},\ #2 (#3)}
\def\prd  #1 #2 #3 {Phys.~Rev.~D~{\bf#1},\ #2 (#3)}
\def\prl  #1 #2 #3 {Phys.~Rev.~Lett.~{\bf#1},\ #2 (#3)}
\def\RMP  #1 #2 #3 {Rev.~Mod.~Phys.~{\bf#1},\ #2 (#3)}
\def\zpc  #1 #2 #3 {Z.~Phys.~C~{\bf#1},\ #2 (#3)}
\def\nim  #1 #2 #3 {Nucl.~Instrum.~Methods~{\bf#1},\ #2 (#3)}
\def\nima  #1 #2 #3 {Nucl.~Instrum.~Methods~A.{\bf#1},\ #2 (#3)}
\def\epjc #1 #2 #3 {Euro.~Phys.~Jour~{\bf#1},\ #2 (#3)}
\def\rmp #1 #2 #3 {Rev.~Mod.~Phys~{\bf#1},\ #2 (#3)}
\def\npbps #1 #2 #3 {Nucl.~Phys.~B.~proc.~suppl~{\bf#1},\ #2 (#3)}
\def\progtp #1 #2 #3 {Prog.~Theo.~Phys~{\bf#1},\ #2 (#3)}
\def\etal{{\it et al.}}

\long\def\inst#1{\par\nobreak\kern 4pt\nobreak
    {\it #1}\par\vskip 10pt plus 3pt minus 3pt}

\begin{document}
{\pagestyle{empty}


\begin{flushright}
\babar-CONF-\BABARPubYear/\BABARConfNumber \\
SLAC-PUB-\SLACPubNumber \\
July 2002 \\
\end{flushright}

\par\vskip 5cm

\begin{center}
\Large \bf Search for the Exclusive Radiative Decays
{\boldmath $\brg$} and {\boldmath $\bomg$} 
\end{center}
\bigskip

\begin{center}
\large The \babar\ Collaboration\\
\mbox{ }\\
July 24, 2002
\end{center}
\bigskip \bigskip

\begin{center}
\large \bf Abstract
\end{center}
A search for the exclusive radiative decays 
$B\to\rho(770)\gamma$ and $\Bz\to\omega(782)\gamma$ 
is performed on a sample of  
\lumi\ $\BB$ events collected by the $\babar$ detector at 
the $\pep2$ asymmetric $\epem$ collider.
No significant signal is seen in any of the channels. We set
preliminary upper limits of $\BR[\brog]< \limitrhoo$, 
$\BR[\brpg]<\limitrhop$ and $\BR[\bomg]< \limitomega$ at 
$90\%$ Confidence Level.
Combining these into a single limit on the generic process \brg,
we find the preliminary limit $\BR[\brg]<\limitgeneric$ corresponding, 
to a limit of \BR[\brg]/\BR[\bkg] $<$ \limitratio\ at $90\%$ Confidence Level.

\vfill
\begin{center}
Contributed to the 31$^{st}$ International Conference on High Energy Physics,\\ 
7/24---7/31/2002, Amsterdam, The Netherlands
\end{center}

\vspace{1.0cm}
\begin{center}
{\em Stanford Linear Accelerator Center, Stanford University, 
Stanford, CA 94309} \\ \vspace{0.1cm}\hrule\vspace{0.1cm}
Work supported in part by Department of Energy contract DE-AC03-76SF00515.
\end{center}

\newpage
} 

\begin{center}
\small

The \babar\ Collaboration,
\bigskip

B.~Aubert,
D.~Boutigny,
J.-M.~Gaillard,
A.~Hicheur,
Y.~Karyotakis,
J.~P.~Lees,
P.~Robbe,
V.~Tisserand,
A.~Zghiche
\inst{Laboratoire de Physique des Particules, F-74941 Annecy-le-Vieux, France }
A.~Palano,
A.~Pompili
\inst{Universit\`a di Bari, Dipartimento di Fisica and INFN, I-70126 Bari, Italy }
% ...

\end{center}\newpage

\section{Introduction}
\label{sec:Introduction}
The effective flavor-changing neutral current processes $\brg$ and $\bomg$ 
probe physics at high mass scales both within the Standard 
Model and within the context of possible new physics scenarios through
the underlying $\bdg$ ``penguin'' transition \cite{hewett}. The decays are 
analogous to the $\bkg$ process mediated by the $\bsg$ transition. The expected
rate of $\bdg$ transitions is suppressed by the ratio of CKM matrix 
elements $|V_{td}/V_{ts}|^2$ relative to $\bsg$ transitions.
There has been considerable interest recently in these exclusive channels, 
resulting in several calculations of the branching fractions expected
in the Standard Model, which indicate a range 
$\BR[\brpg]=(0.9-1.5)\times 10^{-6}$ \cite{SM}.
Though the theoretical uncertainties for the branching fractions remain large,
the possibility of extracting the ratio of CKM elements $|V_{td}/V_{ts}|^2$ 
through the ratio $\BR[B\to(\rho/\omega)\gamma]/\BR[\bkg]$ with 
less uncertainty has been explored 
\cite{alivtdvtstheory}\cite{grinvtdvtstheory}. 
The observation of 
$\brg$ and $\bomg$ would constitute the first evidence of the $\bdg$ 
radiative transition and is of considerable interest as the first step towards 
extracting $|V_{td}/V_{ts}|$ from measurements of these channels. Previous 
searches have
found no evidence for these decays \cite{cleobellerg}.

\section{The \babar\ Detector and Dataset}
\label{sec:babar}
The decay $\brg$ is reconstructed in the modes $\brog$ with 
$\rho^0\to\pip\pim$ 
and $\brpg$ with $\rho^+\to\pip\piz$ 
(charge-conjugate modes are implied throughout), 
while $\bomg$ is reconstructed with $\omega\to\pip\pim\piz$.
The analysis uses a sample of \lumi\ $\BB$ events in $\onlumi\invfb$ 
of data collected by the
$\babar$ detector \cite{ref:detector} 
at the $\pep2$ collider \cite{pep} on the 
$\FourS$ resonance (``on-resonance''), and $9.6\invfb$ of data taken 
$40\mevcc$ below the $\FourS$ resonance (``off-resonance'').   The
reconstruction uses quantities both in the laboratory and 
$\FourS$ center-of-mass (CMS) frames,
where the latter are denoted by an asterisk. The detector 
response to the signal and background processes are studied with a detailed 
Monte Carlo simulation based on Geant4 \cite{geant} and cross-checked 
with control samples in the data. 
The off-resonance data provide a 
control sample of the primary backgrounds from the continuum production
of $u$, $d$, $s$, and $c$ quark-antiquark pairs, while 
exclusively reconstructed $\B\to D\pi$ decays provide a sample
to cross-check the simulation of $\BB$ events.

\section{Analysis Method}
\label{sec:Analysis}

\subsection{Reconstruction of the  Primary Photon}
The primary photon in the decay is identified as a local
maximum within a contiguous deposition of energy in the crystal array of the
electromagnetic  calorimeter (EMC). 
We require that the photon lies in the 
calorimeter acceptance of $-0.74 < \cos\theta< 0.93$,
where $\theta$ is the polar angle to the detector axis.
The energy of the photon, measured in the center of mass system, must satisfy
$1.5 < E_\gamma^* < 3.5$GeV.
The photon candidate is required to be isolated from all other 
local maxima in the calorimeter by 25 cm.
It must also be inconsistent with the trajectories of all reconstructed 
charged tracks. 
Photons consistent with $\piz$ and $\eta$ production are vetoed
by removing candidates that form an invariant mass within $20(40)\mevcc$ of
the $\piz(\eta)$  mass when paired with any other photon in the event with 
energy greater than $50(250)\mev$.  
Energetic $\piz$s which produce photons 
that cannot be resolved as separate local maxima are suppressed by
requiring the lateral profile of the energy deposition to be consistent
with a single photon. 

\subsection{Reconstruction of Charged Tracks}
\begin{figure}[t]
\begin{center}
\includegraphics[height=7.5 cm]{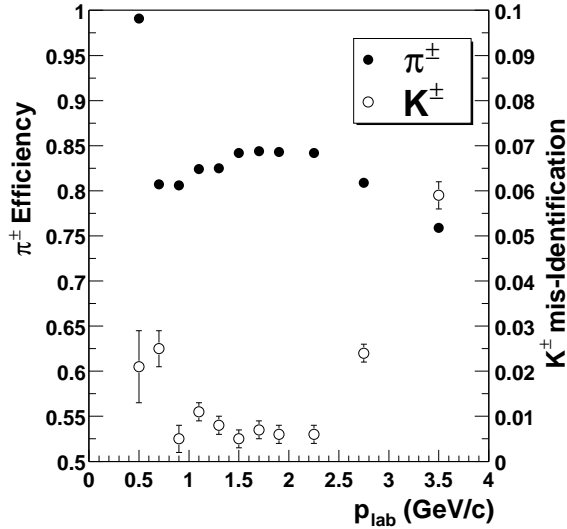}
\caption{\label{fig:pid}Pion selection performance as measured
in data  using $D^{*+} \to D^{0}\pip,D^{0} \to \Km\pip$ decays as a
function of momentum in the laboratory frame $p_{lab}$.
}
\end{center}
\end{figure}

The charged tracks used in identifying the 
$\rho/\omega$ meson are reconstructed in the silicon vertex detector 
(SVT) and drift chamber 
(DCH) and are required to have a trajectory consistent 
with production near the beam interaction point as well as a minimum of
12 hits in the drift chamber. 
A charged pion selection based on $dE/dx$  
measurements and Cherenkov photons reconstructed in the ring-imaging 
Cherenkov detector (DIRC) is used to reduce backgrounds
from $\bkg$ and other $\bsg$ processes by vetoing charged
kaons from these processes.
Both the reconstructed Cherenkov angle and the number of Cherenkov photons observed
is required to be consistent with the pion hypothesis.
Figure \ref{fig:pid} shows the particle identification performance
achieved.
\newline

\subsection{Reconstruction of {\boldmath $\rho/\omega$} Meson}
The $\rho^0\to\pip\pim$ candidates are 
reconstructed by calculating a common vertex for two tracks of opposite charge.
For the $\rho^+\to\pip\piz$ and $\omega\to\pip\pim\piz$ reconstruction, $\piz$ 
candidates are identified as two photon candidates reconstructed 
in the calorimeter each of with energy greater than $50\mev$. 
The invariant mass of the pair is required to be
$0.115 < M_{\gamma\gamma} < 0.150\gevcc$. 
A kinematic fit with 
$M_{\gamma\gamma}$ constrained to the nominal $\piz$ mass is used to improve 
the momentum resolution.  
The $\rho^+$ candidates result from $\piz$ 
candidates paired with an identified charged pion.  
The invariant mass $M_{\pi\pi}$ of the 
$\rho$ candidates is required to be between $0.520$ and $1.020\gevcc$.
The $\omega$ candidates are reconstructed from combinations of 
oppositely charged identified pions 
with a successfully calculated vertex and $\piz$ 
candidates with invariant mass $0.7596 < M_{\pip\pim\piz} < 0.8056\gevcc$.
The momentum of the candidate $\rho$ mesons,
measured in the center of mass system must satisfy
$2.3<p_{\rho}^*<2.85$ and the $\omega$ mesons
must satisfy $2.4<p_{\omega}^*<2.8$.
This cut, which has very high signal efficiency, is
applied in order to reduce the number of events where
more than one candidate satisfies all of the cuts.

\subsection{Reconstruction of the {\boldmath $B$} Meson}
The photon and $\rho/\omega$ meson candidate are 
combined to form the $B$ meson candidate.
The kinematic properties of the $B$ meson are evaluated in the CMS
 using the variables $\de = E^*_{B}-E_{beam}^*$ 
and the beam-energy substituted mass 
$\mes= \sqrt{ E^{*2}_{beam}-\mbox{$\boldmath{\mathrm p'}$}_{B}^{*2}},$
where $E_{beam}^*$ is the energy of the beam, 
$E^*_B=E^*_{\gamma}+E^*_{\rho/\omega}$ is the reconstructed energy 
of the $B$ meson candidate and ${\mathrm p'}_B^*$ is its momentum.
For the purposes of the $\mes$ calculation, 
${\mathrm p'}_B^*$ is modified by scaling the photon energy so as to make
$E^*_\gamma+E^*_{\rho/\omega}-E_{beam}^*=0$, under the assumption that
the resolution of the primary photon energy dominates the $\de$ resolution. 
This procedure reduces the tail in the $\mes$ resolution that
results from the asymmetric photon energy response in the EMC.
The signal events 
have $\mes=m_B$ and  $\de=0$ up to the experimental resolution  of 
$\sim 3\mevcc$ dominated 
by the beam-energy spread in the former case, and $\sim 50 \mev$ dominated
by the reconstructed photon energy resolution in the latter.

We consider candidates in the region $-0.3 < \de <0.3 \gev$ and 
$5.2 < \mes<  5.29\gevcc$ and define a signal region 
of $-0.2 < \de <0.1\gev$ and $5.27<\mes<5.29\gevcc$. Selection criteria have been optimized for best 
$S^2/(S+B)$, where $S$ and $B$ are the expected signal and background yield
assuming $\BR[\brog]=\BR[\bomg]=\frac{1}{2}\BR[\brpg]=10^{-6}$ (as expected
from isospin symmetry) without 
knowledge of the yield or distribution of events in the signal region. 
The signal region extends lower on the negative side of $\de$ due to the 
asymmetric photon energy response in the calorimeter resulting
from energy leakage.
For the small fraction of events (2.8 \% for \brog\ signal)
in which more than one $B$ meson candidate satisfies all the cuts, 
the candidate with the smallest value of $|\de|$ is selected.

\begin{figure}[t]
\begin{center}
\includegraphics[height=7.5 cm]{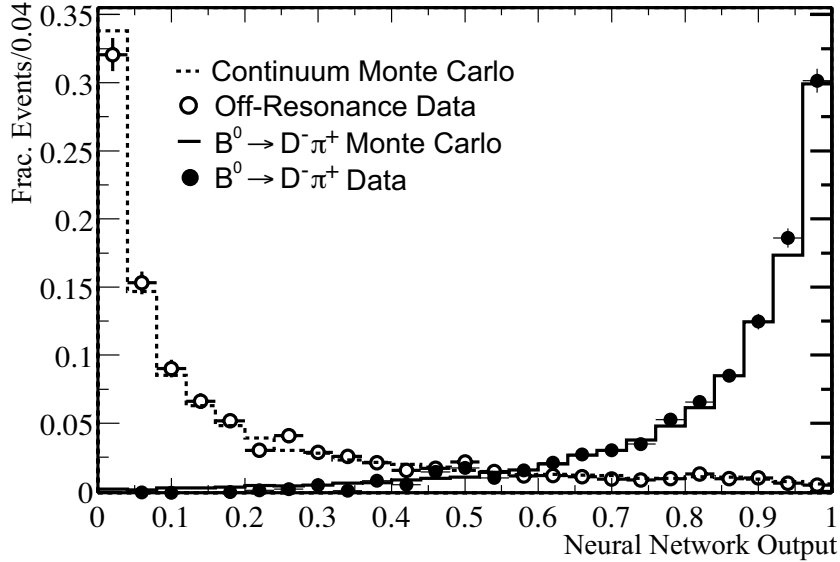}
\caption{ $\brog$ neural network output for Monte
Carlo-simulated events with comparison to  data control samples.} 
\label{fig:rhonn}
\end{center}
\end{figure}

\subsection{Suppression of Background}
The continuum and initial state radiation backgrounds are 
suppressed by a neural network that combines event topology variables into
one discriminating variable \cite{snns}. The neural network responds 
non-linearly
to the input variables and exploits correlations between the variables.
The input variables are:
\begin{itemize}
\item{$\ctt$, the 
cosine of the angle between the photon and the thrust axis of 
the rest of the event (excluding the $B$ meson candidate).}
\item{$\ctb$, the cosine of the angle between
the $B$ meson momentum and the beam axis.}
\item{The energy flow in $10^\circ$ bins centered on the 
photon candidate momentum in the CMS.}
\item{$\cth$, the cosine of the helicity angle.
For $\brg$, we define $\theta_H$ as the angle between the $\pip$ momentum
in the $\rho$ rest frame and the $\rho$ momentum in the $B$ meson rest frame. 
For $\bomg$,  $\theta_H$ is defined as the angle between the normal to 
the plane defined by the $\pip\pim\piz$ momenta in the $\omega$ rest frame and
the $\omega$ momentum in the $B$ meson rest frame.
$\cth$ should follow a  $\sin^2\theta_H$ distribution for signal, while 
the continuum background is approximately flat. }
\item{$R_2'$, the ratio of second and zeroth order Fox-Wolfram moments 
in the frame recoiling from the photon momentum.
This is effective against initial-state radiation, since 
in that frame the jet structure of 
the hadrons is recovered.}
\item{The net flavor content, defined as 
$\sum_i |N^+_i-N^-_i|$, where $N^\pm_i$ are the number of $e^\pm$, 
$\mu^{\pm}$, $K^\pm$ and slow pions of each sign identified in the event.}
\item{$|\Delta z|$,  the vertex separation of the $B$ meson candidate and 
the rest of the event along the beam axis, is used for 
$\brog$ and $\bomg$.}
\item{$\cos\theta_D$, the cosine of the Dalitz 
angle of the $\omega$ decay is used for $\bomg$.
$\cos\theta_D$ is defined as the angle between the.
$\piz$ and the $\pip$ momenta
in the rest frame of the $\pi^+\pi^-$ system.
We expect $\cos\theta_D$ to be uniformly distributed for
the combinatorial background and to follow a
$\sin^2\theta_D$ distribution for true $\omega$ decays.}
\end{itemize}

A separate neural network is trained for each mode using the  
back-propagation algorithm on samples of Monte Carlo-simulated 
signal and background. The output of the neural network, defined such
that the signal processes peak at one and the continuum background at zero,
is cross-checked on an independent sample of Monte 
Carlo-simulated events and data control samples for both the signal and 
background.
The neural network output for $\brog$ is shown in 
Figure~\ref{fig:rhonn}, where the Monte Carlo simulation of the continuum
background is compared with the off-resonance data and the output for
Monte Carlo-simulated $\Bz\to D^-\pip$ decays is compared with events
reconstructed in the data. 
This latter check gives us confidence that the Monte Carlo simulates
the $\brg$ efficiency well, since most of the input variables of the neural 
network are not based on the properties of the signal decay itself, but rather
on the properties of the other $B$ meson in the event. 

We make  a selection on the neural network output optimized for best 
$S^2/(S+B)$ for each mode. For $\brpg$, an additional requirement of
$|\cth|<0.6$ is made to reject $\Bu\to\rho^+\piz$ events which have 
a $\cos^2\theta_H$ distribution, different from the expected $\sin^2\theta_H$ 
distribution from the signal process.

\begin{figure}[t]
\begin{center}
\includegraphics[width=10 cm]{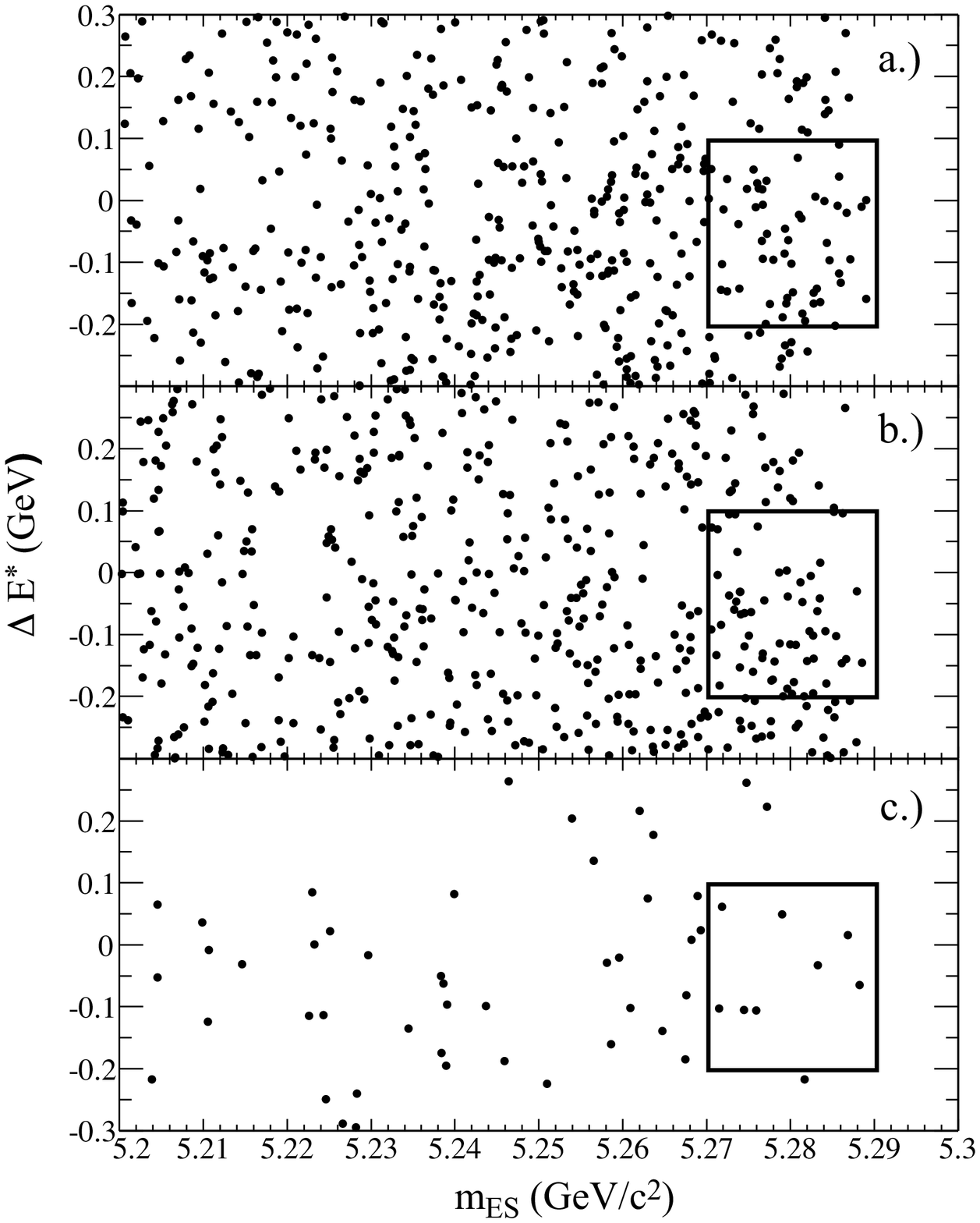} 
\caption{$\de$ vs. $\mes$ for a.) $\brog$, b.) $\brpg$ and c.) $\bomg$ candidates.}
\label{fig:rhomesde}
\end{center}
\end{figure}

\subsection{Background Estimation}
Table \ref{tab:background} shows estimates for the expected 
background remaining in the signal region
($-0.2 < \de < 0.1 \gev$ and $\mes>5.27\gevcc$) 
after the neural network
selection, using the off-resonance data for continuum and Monte Carlo 
simulation for $\BB$ backgrounds.
There is good agreement between Monte Carlo estimates of continuum
background and off-resonance data.
The $\BB$ background is smaller than the continuum background. 

\begin{table}[tb]
\begin{center}
\begin{tabular}{lccc} \hline \hline
&$\brog$&$\brpg$&$\bomg$\\
&\multicolumn{3}{c}{(Events)} \\\hline
Off-resonance data& 
$42.8 \pm 7.0$ & $66.5 \pm 8.8$ &$7.3 \pm 3.3$\cr
$uds+c\bar c+\tau^+\tau^-$ Monte Carlo& 
$39.8 \pm 3.8$ & $59.8 \pm 4.6$ & $8.4 \pm 1.9$\cr
\hline 
$\BB$ Monte Carlo & 
$\;\> 6.6 \pm 1.0$ & $12.2 \pm 3.6$&$0.7 \pm 0.3$\cr
\hline
Signal Expectation & 
$\;\> 9.9 \pm 0.2$&$12.1 \pm 0.3$&$3.4 \pm 0.1$\cr
\hline\hline
\end{tabular}
\caption{\label{tab:background} Estimates of continuum background
(using both off-resonance data and Monte Carlo simulation), $\BB$ backgrounds
and signal in the signal region.
All estimates are scaled to a luminosity corresponding to \lumi\ 
$\BB$ pairs.
Errors are statistical only.}
\end{center}
\end{table}

\subsection{Signal Extraction}
After the neural network selection,
the signal extraction in $\brg$ is performed with an unbinned extended maximum 
likelihood fit in the variables $\mes$, $\de$ and $M_{\pi\pi}$ with signal
and continuum background components. 
Since the $\BB$ backgrounds are small relative to the continuum background,
the signal extraction uses only a continuum component 
to describe the background.
Biases due to $\BB$  backgrounds are estimated in the systematic studies
described in Section~\ref{sec:Systematics}.
The signal $\mes$ and $\de$ distributions 
are described by the Crystal Ball lineshape \cite{cbshape}, with the exception
of the $\mes$ distribution for $\brog$, where the Gaussian distribution is 
used. Here, we expect the photon energy rescaling to eliminate the tail. The 
background $\mes$ and $\de$ distributions are
described by the ARGUS threshold function \cite{argus} and second order polynomial,
respectively. For $\brg$, the Breit-Wigner lineshape is used for the
signal $M_{\pi\pi}$ distribution, while the background incorporates a sum
of a resonant Breit-Wigner component and a continuum component described by a 
first order polynomial.

The parameters of the signal
probability distributions are obtained from the Monte Carlo simulation and 
cross-checked in the data with the decays $\bkog$, $\Kstarz\to\Kp\pim$ 
for $\brog$ and  $\bkpg$, $\Kstarp\to\Kp\piz$ for $\brpg$, which are 
topologically and kinematically similar. The
parameters of the continuum background distributions are determined
in the fit, with the exception of the fraction of the resonant 
$\rho\to\pi\pi$ contribution to the continuum $M_{\pi\pi}$ distribution,
which is fixed to the value measured in the off-resonance data. 
In the $\bomg$, the signal extraction is performed in a similar fit 
to the $\mes$ and $\de$ distributions; 
the $M_{\pip\pim\piz}$ distribution is not included
in the fit, because the continuum background contains a large 
and uncertain fraction of true $\omega$ decays, which is difficult to model.

By inverting the pion selection on the charged pion (and selecting kaons)
in the $\brpg$ analysis and on one of the charged pions in 
the $\brog$ analysis, we obtain an 
orthogonal sample of events enhanced in the decays $\bkpg$, 
$\Kstarp\to\Kp\piz$ and $\bkog$, $\Kstarz\to\Kp\pim$, respectively. 
The yield 
of $\bkg$ in these two samples is determined using the same fit procedure 
described for $\brg$, with the expected signal distributions 
determined from Monte Carlo simulation. 
The resulting yields are in agreement 
with the expectations from previous measurements of $\BR[\bkg]$ 
\cite{babarksg}, as shown in Table~\ref{tab:ksgyield}, thus providing a 
cross-check on the event selection 
and signal extraction procedure up to the statistical uncertainty in the 
extracted yields and the uncertainties in the measured branching fractions. 

The $\de$ vs. $\mes$ distributions of the $\brg$ and $\bomg$ candidates are shown
in Figure~\ref{fig:rhomesde} and the fitted yields summarized in
Table~\ref{tab:sigyield}. 
The quality of the fit is determined  by 
comparing the minimum $-\log\mathcal {L}$ of the fit, 
where $\mathcal{L}$ is the
overall likelihood of the fit,  with values
obtained from parameterized Monte Carlo simulation and found to be in good 
agreement.

\begin{table}
\begin{center}

\begin{tabular}{lcc} \hline\hline
Mode		        & Fitted Yield (Events)         & Expected Yield (Events) \\  \hline
$\bkog$                 & $343.2 \pm 21.0$              & $332\pm 36$\\ 
$\bkpg$                 & $\;\>93.1\pm12.6$               	& $105\pm 18$\\  \hline\hline
\end{tabular}
\caption{ \label{tab:ksgyield} The fitted yields in the $\bkg$-enhanced sample described 
in the text, and the expected yield from the measured branching fractions in \cite{babarksg}.}

\vskip 0.3 cm

\begin{tabular}{lcccc} \hline\hline
Mode		& Yield		& $90\%$ C.L. Upper Limit Yield 	& Bias		& Efficiency	\\ 
		& (Events)	& (Events)		& (Events)	& $(\%)$	\\ \hline
$\brog$		& $4.8\pm5.2$	& 12.4			& [-0.5,0.8] 	&	12.3	\\
$\brpg$		& $6.2\pm5.5$	& 15.4			& [-0.1,2.0]	&$\>$ 9.2	\\
$\bomg$		& $0.1\pm2.3$	& $\>\;$3.6		& [-0.3,0.5]	&$\>$ 4.6\\ \hline\hline 
\end{tabular}
\caption{ \label{tab:sigyield} 
The fitted yields, the ranges of observed biases from $\BB$ backgrounds 
and selection
efficiencies for $\brog$, $\brpg$ and $\bomg$ in the on-resonance data sample.
The efficiencies include the branching fractions for 
$\omega\to\pip\pim\piz$ (88.8\%), $\rho^0\to\pip\pim$ ($\approx$ 99\%)
and $\rho^+\to\pip\piz$ ($\approx$ 100\%).}
\end{center}
\end{table}

\section{Systematic Studies}
\label{sec:Systematics}
The systematic uncertainties in this analysis are associated with uncertainties
in the efficiency of the signal process reconstruction predicted by the
Monte Carlo simulation and with the signal extraction procedure. 
Table \ref{tab:rhosystab} summarizes these uncertainties.
The efficiency of the track selection is calculated by 
identifying tracks in the silicon vertex detector and evaluating the fraction 
that is well-reconstructed in the drift chamber. 
The pion identification efficiency in the DIRC is derived from a sample of 
$D^{*+} \to D^{0}\pip,D^{0} \to \Km\pip$ decays. 
The photon and $\piz$ efficiencies are measured
by comparing the ratio of events  
\mbox{$N(\tau^{\pm}\to h^{\pm}\piz)/N(\tau^{\pm}\to h^{\pm}\piz\piz)$}, 
where $h^\pm=\pi^\pm,~K^\pm$, 
to the previously measured branching ratios \cite{cleo3}. 
The photon isolation and $\piz/\eta$ veto efficiency
are dependent on the event multiplicity and are tested by ``embedding'' 
Monte Carlo-generated photons into both an exclusively reconstructed $B$ meson
data sample  and a generic $\BB$ Monte Carlo sample. The efficiencies of the
neural network selection are compared between the exclusively reconstructed 
$B\to D\pi$ events in the data and the Monte Carlo simulation of both
the $B\to D\pi$ and signal processes, and the observed variations taken as
systematic uncertainties. 
The $M_{\pip\pim\piz}$ selection for the $\bomg$ is checked
by comparing the resolution of the $\omega$ mass peak in the 
on-resonance data with the Monte Carlo simulation and evaluating the variation
in efficiency.

Systematic uncertainties in the signal extraction procedure result from the 
modeling of the $\BB$ backgrounds and from uncertainties in the signal 
probability distribution functions used in the fit. 
The biases due to $\BB$ backgrounds
are estimated by varying the distributions as well as the rates
of the dominant backgrounds coming from $\bsg$ and $\Bu\to\rho^+\piz$ decays. 
The full range of biases obtained from these variations is taken as the
allowed range, as shown in 
the column labeled ``Bias'' in Table \ref{tab:sigyield}.
The uncertainties resulting from
the fixed parameters describing the
signal distributions in the fit
are estimated by varying the parameters within the uncertainty 
obtained from the analogous $\bkg$ 
processes in the data used to cross-check the expectations from the
Monte Carlo simulation. The effects of these variations on the fitted signal 
yield in each mode is calculated and the range of observed biases are taken
as systematic uncertainties in the signal yield. 

\begin{table}[tbh]
\begin{center}
\begin{tabular}{lccc} \hline \hline
			& $\brog$       & $\brpg$  & $\bomg$ 	\\  
& \multicolumn{3}{c}{Systematic Uncertainty} \\  
Selection Criteria      & $(\%)$        & $(\%)$   & $(\%)$	\\ \hline
$B$ Count               & 1.1           & 1.1      & 1.1 	\\
$\gamma$ Eff.           & 1.5           & 1.5      & 1.5 	\\
$\piz$ Eff.             & -	        & 5.0      & 5.0 	\\
$\piz/\eta$ Veto        & 1.0           & 1.0      & 1.0 	\\
$\gamma$ Dist Cut       & 2.0           & 2.0      & 2.0 	\\
Tracking Eff.           & 2.5           & 1.3      & 2.4 	\\
$\pi$ Selection         & 6.0           & 3.0      & 6.0 	\\ 
$M_{\pip\pim\piz}$ Selection & -        & -        & 2.0 	\\ 
Neural network Selection     & 8.0      & 6.0      & 14.0	\\
Fit Distributions       & 5.0           & 10.0     & 5.0	\\ \hline
Total                   & $11.8\%$      & $13.4\%$ & $17.3\%$\\ \hline \hline
\end{tabular}
\caption{\label{tab:rhosystab} Summary of systematic uncertainties for $\brg$ and 
$\bomg$ expressed as percent error of the branching fraction. }
\end{center}
\end{table}

\section{Physics Results}
\label{sec:Physics}
Based on the observed signal yields we determine $90\%$ upper limits on the
branching fraction by re-fitting the distributions with increasing fixed signal
yields until the $-\log\mathcal{L}$ deviates by 0.82 relative to the minimum
value. We correct for bias from $\BB$ backgrounds by applying the smallest 
observed bias to the signal yield (increasing the signal 
yield in all cases). 
The yield of events in the data sample are calculated with the 
Monte Carlo-derived efficiency lowered by one standard deviation in the 
systematic error. Finally, the estimated number of $\BB$ events in the
sample is reduced by one standard deviation. 
The resulting preliminary $90\%$ confidence
level upper limits for the branching fractions are
$\BR[\brog]<\limitrhoo$, $\BR[\brpg]<\limitrhop$ and
$\BR[\bomg]<\limitomega$.

For the purpose of comparing the limits to $\BR[\bkg]$, we 
combine these limits into a single limit on the generic process
$\brg$ defined as
$$
\BR[\brg] \equiv \BR[\brpg] = 2\times\BR[\brog] = 2\times\BR[\bomg].
$$
as expected from isospin symmetry.
The resulting preliminary $90\%$ confidence level upper limit is
\newline
$\BR[\brg] < \limitgeneric$.
Using the measured value of \BR[\bkg] \cite{babarksg}, this corresponds to a ratio of 
$$
\BR[\brg]/\BR[\bkg] < \limitratio.
$$

We can convert this ratio to a limit on $|V_{td}/V_{ts}|$
using the following expression from  \cite{alivtdvtstheory}.
\[
\frac{{\cal B}[\brg]}{{\cal B}[\bkg]}=
\left| \frac{V_{td}}{V_{ts}} \right|^{2}
\left(\frac{1-m_{\rho}^{2}/M_{B}^{2}}{1-m_{K^{*}}^{2}/M_{B}^{2}}\right)^{3}
\zeta^{2} [1+\Delta R].
\]
Taking the conservative limits of the two parameters, $\zeta = 0.7$ and $\Delta R = -0.25$,
we find 
\newline
$|V_{td}/V_{ts}| < 0.36$ at $90\%$ Confidence Level.

\section{Summary}
\label{sec:Summary}
In conclusion, we have found no evidence for the exclusive $\bdg$ transitions
$\brg$ and $\bomg$ in \lumi\ $\BB$ decays studied with the $\babar$ detector.
The preliminary $90\%$ confidence level upper limits on the branching 
fractions are
significantly improved and within a factor of two of the largest Standard 
Model predictions, which indicate a range $\BR[\brpg]=(0.9-1.5)\times 10^{-6}$.

\section{Acknowledgments}
\label{sec:Acknowledgments}


We are grateful for the 
extraordinary contributions of our \pep2\ colleagues in
achieving the excellent luminosity and machine conditions
that have made this work possible.
The success of this project also relies critically on the 
expertise and dedication of the computing organizations that 
support \babar.
The collaborating institutions wish to thank 
SLAC for its support and the kind hospitality extended to them. 
This work is supported by the
US Department of Energy
and National Science Foundation, the
Natural Sciences and Engineering Research Council (Canada),
Institute of High Energy Physics (China), the
Commissariat \`a l'Energie Atomique and
Institut National de Physique Nucl\'eaire et de Physique des Particules
(France), the
Bundesministerium f\"ur Bildung und Forschung and
Deutsche Forschungsgemeinschaft
(Germany), the
Istituto Nazionale di Fisica Nucleare (Italy),
the Research Council of Norway, the
Ministry of Science and Technology of the Russian Federation, and the
Particle Physics and Astronomy Research Council (United Kingdom). 
Individuals have received support from 
the A. P. Sloan Foundation, 
the Research Corporation,
and the Alexander von Humboldt Foundation.


\begin{thebibliography}{99}
\bibitem{cbshape}
The ``Crystal Ball'' lineshape is a modified Gaussian
distribution with a transition to a tail function on one side:
$ f_{CB} = \exp{-\frac{(x-\mu)^2}{2\sigma^2}}$ for 
$\frac{x-\mu}{\sigma} > \alpha$ and $A\times[B-(\frac{x-\mu}{\sigma})^{-n}]$ 
for $\frac{x-\mu}{\sigma} < \alpha$ where
$ A \equiv \left(\frac{n}{\alpha}\right)^n\exp{-\frac{1}{2}|\alpha|^2}$
and $ B \equiv \frac{n}{\alpha}-|\alpha|$ are defined such as to maintain
continuity of the function and its first derivative.
\bibitem{argus}
We use the distribution $x\sqrt{1-x^2}\times\exp[\zeta(1- x^2)]$,
where $x=\mes/E^*_{beam}$, to describe the background $\mes$ distribution.
The function was introduced by the ARGUS collaboration; 
H.Albrecht \etal, \zpc 48 543 1990
\end{thebibliography}
\end{document}